# AI-Automation Tooling in Computer Engineering Education: Mixed-Methods TAM/UTAUT Evidence for a General Acceptance Attitude

Aung Pyae
*International School of Engineering,*
*Faculty of Engineering*
*Chulalongkorn University, Thailand*
aung.p@chula.ac.th

***Abstract*—As generative AI and low-code workflow platforms become routine in software practice, a key educational question is whether the next generation of computer engineers will accept these tools as useful, usable, and worth sustained engagement. This paper reports a mixed-methods, cross-sectional study of undergraduate computer engineering students' acceptance of AI-automation tooling—instantiated via the open-source platform n8n across three identically scripted workshops in Thailand (n = 103). A 12-item, five-point Likert instrument mapped to six TAM/UTAUT constructs—Performance Expectancy (PE), Effort Expectancy (EE), Behavioral Intention (BI), Self-Efficacy (SE), Hedonic Motivation (HM), and Output Quality (OQ)—was complemented by inductive thematic analysis of open-ended feedback. Analyses combined ordinal reliability estimation, bootstrap confidence intervals, non-parametric tests, multiple-comparison-controlled correlations, polychoric dimensionality diagnostics, a common-method-bias check, and between-session comparisons. Acceptance was favourable across all six constructs at large effect sizes—PE strongest, HM weakest—and dimensionality diagnostics revealed that canonical TAM/UTAUT sub-facets collapse into a single general acceptance factor in this short-form post-workshop context, a methodologically and theoretically consequential finding. Qualitative themes converged with the quantitative profile on usefulness and enthusiasm but diverged on output quality, surfacing a small yet articulate reliability-sceptical minority. Findings support the curricular adoption of AI-automation tooling in undergraduate computing education and identify three theory-grounded instructional levers: instruction-sequencing scaffolds, self-efficacy supports, and trust-calibration interventions.**

***Keywords—technology acceptance model; UTAUT; AI automation; low-code platforms; student attitudes; self-efficacy; output quality; trust in automation***

## I. INTRODUCTION

Generative artificial intelligence (AI) and low-code workflow automation platforms are reshaping how engineering students conceive of and construct software. As the next generation of software professionals will be expected to design, deploy, and reason about AI-augmented automations in their working life, an open empirical question for computer engineering education is straightforward: do undergraduate students actually accept these tools, and what attitudes do they bring to them, when the tools are introduced through controlled instructional workshops?

This paper reports a mixed-methods, cross-sectional study of that question among Thai computer engineering undergraduates. Acceptance and attitudes were measured immediately after three identically scripted workshops in which AI-automation tooling was instantiated through the open-source platform n8n [1]. n8n is the instantiation, not the topic: the contribution of the paper is what the data reveal about students' acceptance of, and attitudes toward, AI-automation tooling in education, not an evaluation of the platform itself.

The work is grounded in the Technology Acceptance Model (TAM) [2], the Unified Theory of Acceptance and Use of Technology (UTAUT) [3], and its consumer extension UTAUT2 [4]. Of 120 invited participants, 103 voluntarily consented and completed a 12-item Likert questionnaire mapped to six TAM/UTAUT constructs: Performance Expectancy (PE), Effort Expectancy (EE), Behavioral Intention (BI), Self-Efficacy (SE), Hedonic Motivation (HM), and Output Quality (OQ). Quantitative item- and construct-level statistics were complemented by an inductive thematic analysis of open-ended student feedback. Because the measurement design is deliberately parsimonious (two items per construct), the conventional descriptive analyses are augmented with polychoric-based dimensionality diagnostics, ordinal reliability (McDonald's omega), bootstrap confidence intervals, multiple-comparison control, and a common-method-bias assessment—procedures that squarely address the principal measurement-validity concerns that such a design would otherwise invite. The study is guided by three research questions:

- RQ1. To what extent do computer engineering undergraduates accept AI-automation tooling across the six TAM/UTAUT constructs, and is their acceptance reliably above the neutral midpoint of the scale?
- RQ2. Which TAM/UTAUT dimensions are strongest and weakest in this student population, and how are the six constructs interrelated—including whether they are empirically separable in this sample?
- RQ3. How do qualitative student narratives converge with or diverge from the quantitative acceptance profile, and what instructional implications follow?

The contribution is fourfold. Empirically, the paper provides construct-level evidence on Thai computer engineering undergraduates' acceptance of AI-automation tooling—a population under-represented in the TAM/UTAUT literature. Methodologically, it offers a transparent, instrument-matched analytical account—polychoric inputs for ordinal items, Spearman–Brown and McDonald's omega for two-item scales, non-parametric midpoint tests, and explicit dimensionality and common-method-bias diagnostics—rather than defaulting to structural equation modelling for which sample size and indicator counts are insufficient. Theoretically, it documents that canonical TAM/UTAUT sub-facets are not empirically separable in short-form post-workshop measurement, behaving instead as sub-dimensions of a general acceptance attitude—with direct implications for how short acceptance instruments should be deployed in computing-education research. Pedagogically, the mixed-methods integration identifies Output Quality as the locus of quantitative–qualitative divergence and produces three actionable instructional levers grounded in the trust-in-automation literature.

## II. Literature Review

### A. Theoretical Foundations: TAM, UTAUT, and UTAUT2

The Technology Acceptance Model (TAM) [2] posits that behavioural intention to use a technology is determined principally by perceived usefulness and perceived ease of use; the longitudinal extension TAM2 [5] identifies output quality and result demonstrability as antecedents of perceived usefulness, and Davis, Bagozzi, and Warshaw [14] establish that intrinsic enjoyment contributes to use beyond utilitarian benefit. The Unified Theory of Acceptance and Use of Technology (UTAUT) [3] consolidates eight competing acceptance models around four determinants—performance expectancy, effort expectancy, social influence, and facilitating conditions—and its consumer extension UTAUT2 [4] introduces hedonic motivation, price value, and habit. The present study adopts performance expectancy, effort expectancy, behavioural intention, and hedonic motivation from the UTAUT family [3], [4], complemented by self-efficacy [6], [7] and output quality [5]. Social influence and facilitating conditions are deliberately excluded because both are largely homogenized by the classroom context (a shared instructor cohort, uniform teaching-assistant support, and identical technical infrastructure across sessions), which would suppress construct variance and confound interpretation with the workshop environment itself.

### B. Self-Efficacy in Technology Acceptance

Bandura's social-cognitive account [6] characterizes self-efficacy—the perceived capability to execute the actions required to attain a given outcome—as a central antecedent of behavioural engagement. Compeau and Higgins [7] operationalized this construct for the computing context and demonstrated computer self-efficacy to be a robust predictor of both anticipated and actual technology use, a relationship reproduced in subsequent acceptance research. In educational settings, self-efficacy is particularly scaffold-sensitive: instructional support, worked examples, and graduated independence directly shape learners' confidence to construct artefacts unaided, motivating its inclusion as a sixth construct alongside the canonical UTAUT variables.

### C. Output Quality, Trust in Automation, and Low-Code Platforms in Computing Education

Output quality is a TAM2 antecedent of perceived usefulness [5] and, for automated systems, is closely intertwined with the trust-in-automation literature. Lee and See [8] frame trust as an attitude that is appropriate when calibrated to the automation's actual capabilities; Hoff and Bashir [9] decompose its empirical determinants into dispositional, situational, and learned layers; Parasuraman and Riley [10] articulate how use, misuse, disuse, and abuse of automation arise when reliance is poorly calibrated to reliability; and Parasuraman, Sheridan, and Wickens [11] supply a stages-and-levels model that locates such reliance within the broader human–automation interaction loop. For AI-augmented workflow platforms—where users delegate multi-step tasks to opaque large-language-model agents—recent human–AI work shows that interface design and cognitive-forcing functions shape whether users verify outputs or defer to the system [12], [13]. Low-code development platforms reduce the textual-programming burden through visual abstractions and reusable connectors [15]; n8n's open-source licence, node-based visual semantics, and native AI-model integration make it a credible candidate for curricular adoption [1], yet empirical acceptance evidence for such platforms among undergraduate engineers remains sparse—a gap this study addresses.

### D. Methodological Considerations and Research Gap

Single-source self-report instruments are susceptible to common-method bias [16], and two-item-per-construct scales can inflate inter-construct correlations and erode discriminant validity. For two-item scales, Spearman–Brown [17] is preferable to Cronbach's alpha [18], with McDonald's omega [19] as a congeneric alternative. Harman's single-factor diagnostic [16], Horn's parallel analysis [20], and the Kaiser–Meyer–Olkin measure [21] together enable principled dimensionality and common-method-bias assessment without a confirmatory factor model. Synthesizing the foregoing, six constructs (PE, EE, BI, SE, HM, OQ) are theoretically apposite for undergraduates' acceptance of AI-automation tooling, predicting (i) above-neutral endorsement in a population already familiar with general-purpose AI tools, (ii) positive inter-construct correlations consistent with the TAM/UTAUT nomological network [2]–[5], and (iii) PE ranking highest given the concrete instrumental utility students perceive for assignments, projects, and careers, while anticipating only modest differentiation among constructs under short-form measurement. The study tests these expectations, uses the qualitative stream to probe attitudinal divergences, and treats dimensionality and common-method bias as empirical questions rather than assumptions.

## III. Method

### A. Design and Participants

A cross-sectional, explanatory mixed-methods design combined a self-administered Likert questionnaire with an optional open-ended item; the design is correlational and no causal inference is intended. Participants were computer engineering undergraduates at a Thai university with basic-to-intermediate Python competence, daily use of general-purpose AI tools (e.g., ChatGPT), and foundational AI/data-science knowledge but no prior AI-development experience. Three workshops (~40 students each) yielded 120 invited participants; 103 consented and completed the questionnaire (response rate 85.83%), with both female and male students represented. Session assignment was reconstructed from timestamped submissions, producing three clusters: Workshop 1, $n = 31$ (30 March 2026); Workshop 2, $n = 45$ (31 March–1 April 2026); Workshop 3, $n = 27$ (16–17 April 2026). This reconstruction is acknowledged as a limitation.

### B. Workshop Protocol

All three workshops followed the controlled protocol summarized in Fig. 1. Sessions were held in a classroom on students' own browser-based devices (n8n required no installation); pre-session account setup (n8n, CodePen, Groq) was completed by participants in advance. The instructor presented n8n's workflow concepts, demonstrated a reference implementation, and supported guided practice and an independent task with the assistance of teaching assistants and a written user manual. An illustrative canvas of the kind built during the sessions is shown in Fig. 2: a Trigger node feeds an AI-Agent node connected to short-term Memory, with results emitted to an Output node. Protocol fidelity was maintained by using the same instructor, manual, reference implementation, task brief, and TA team across all three sessions; total content time was approximately two hours plus an optional ten-minute survey administered after the debrief.

**Three identically scripted workshops**
Subsamples: $n$ = 31 | 45 | 27
Combined $n$ = 103

|  |  |
|---|---|
| PRE-SESSION | Account setup (n8n, Groq, CodePen) |
| PHASE 0 | Introduction ~10 min |
| PHASE 1 | Live demonstration ~20 min |
| PHASE 2 | Guided practice ~60 min |
| PHASE 3 | Independent task ~30 min |
| PHASE 4 | Debrief & Q&A ~10 min |
| PHASE 5 | Survey (voluntary) ~10 min |

*Identical instructor, written manual, reference workflow, and TA team*

Fig. 1. Workshop protocol used in all three identically scripted sessions; the survey was administered after the debrief and was strictly voluntary.

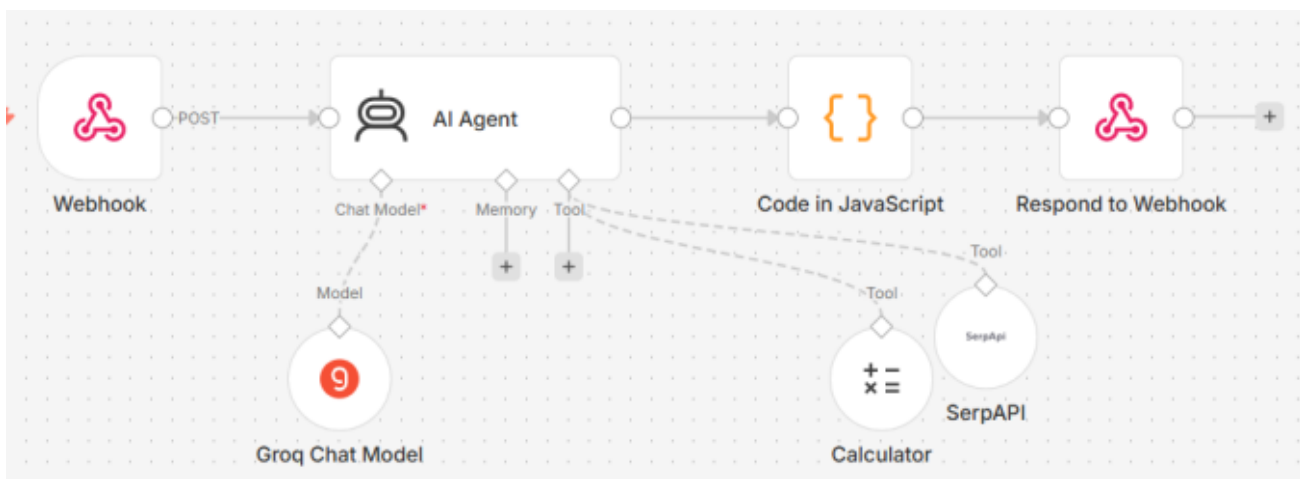

Fig. 2. Illustrative n8n canvas from the workshop showing a simple Trigger → AI Agent → Output flow with attached memory.

## C. Instrument and Data Collection

The 12-item questionnaire used a five-point Likert scale (1 = Strongly Disagree, 5 = Strongly Agree). Items were adapted from the TAM/UTAUT family [2]–[5] and Bandura's self-efficacy theory [6], [7], worded for the AI-automation context, with two items per construct (PE, EE, BI, SE, HM, OQ). The two-item-per-construct design was chosen to minimize respondent burden after a multi-phase workshop; Eisinga et al. [17] specify Spearman–Brown and congeneric omega as the appropriate reliability estimators, both of which are reported. The instrument was administered in fixed item order via Google Forms (no reverse-coded items); an optional free-text item elicited open feedback. Data were collected between 30 March and 17 April 2026 (GMT+7). Consent was obtained explicitly before any items were shown; only records with affirmative consent were retained (n = 103, 100% of submissions). No demographic variables were collected, to minimise identifiability. There were zero missing values across the 103 respondents, and students were informed that non-participation would not affect their course standing.

## D. Data Analysis

Quantitative analyses comprised: (a) item- and construct-level descriptives; (b) construct scores as the mean of two items, with reliability via Cronbach's alpha [18], Spearman–Brown [17], and McDonald's omega [19] from polychoric correlations; (c) bootstrap (B = 5,000) 95% CIs for construct means; (d) Spearman correlations with bootstrap 95% CIs and Holm correction [22] across 15 pairwise tests; (e) one-sample Wilcoxon signed-rank tests against the neutral midpoint 3.0, with effect size r = |Z|/√n_nz and Cohen's benchmarks [23]; (f) dimensionality diagnostics on the 12×12 polychoric matrix—KMO [21], Bartlett's test, Horn's parallel analysis (500 replications) [20], and Harman's single-factor test [16]; (g) between-session Kruskal–Wallis tests; and (h) respondent segmentation into High (≥ 4.00), Moderate (3.00–3.99), and Low (< 3.00) acceptors per Likert-response conventions [24]. Structural equation modelling and confirmatory factor analysis were deliberately excluded: with n = 103 and two indicators per construct, a six-factor model is under-identified. Qualitative analyses followed Braun and Clarke's inductive thematic analysis [25]; substantive comments were coded across two passes separated by 48 hours by a single analyst, retaining only stable themes; inter-rater reliability was not computed and is acknowledged as a limitation.

# IV. Results and Findings

## A. Descriptive Statistics and Reliability

All twelve items recorded means above 3.9 (Q2 [PE] highest at 4.36; Q7 [SE] lowest at 3.91), with the share rating Agree or Strongly Agree (%Agree) ranging from 66.99% to 85.44%. Construct-level means, bootstrap 95% CIs, and three reliability estimates are shown in Table I. All six construct-mean 95% CIs lie entirely within the "agree" region of the scale. McDonald's omega [19] ranges from 0.82 (SE) to 0.94 (BI); Spearman-Brown coefficients and Cronbach's alpha [17], [18] both range from 0.76 to 0.90. All three estimators agree that the six two-item scales show acceptable internal consistency, recognizing that with only two indicators per construct the interpretive range of these coefficients is structurally narrower than for ≥3-item scales. The lowest-reliability construct is SE, consistent with the SE pair tapping partially distinct aspects (confidence to build unaided vs. ability to learn). Ranking by mean gives PE (4.28) > OQ (4.16) > EE (4.11) > BI (4.05) > SE (4.04) > HM (4.00); the full range is only 0.28 scale points, so the ordering is interpretively useful rather than evidence of steep differentiation.

TABLE I. Construct-Level Descriptives, Reliability, and Bootstrap 95% CIs

| Construct | Mean | 95% CI | $\alpha$ | SpB | $\omega$ |
|---|---|---|---|---|---|
| PE — Performance Exp. | 4.28 | [4.11, 4.44] | 0.81 | 0.81 | 0.88 |
| EE — Effort Exp. | 4.11 | [3.91, 4.29] | 0.85 | 0.85 | 0.90 |
| BI — Behavioral Int. | 4.05 | [3.84, 4.25] | 0.90 | 0.90 | 0.94 |
| SE — Self-Efficacy | 4.04 | [3.85, 4.21] | 0.76 | 0.76 | 0.82 |
| HM — Hedonic Mot. | 4.00 | [3.79, 4.19] | 0.85 | 0.85 | 0.89 |
| OQ — Output Quality | 4.16 | [3.96, 4.33] | 0.77 | 0.78 | 0.84 |

## B. One-Sample Tests against the Neutral Midpoint

Table II reports one-sample Wilcoxon signed-rank tests of each construct score against the neutral midpoint 3.0. All six tests were significant at p < 0.0001 and remained significant after Holm correction. Effect sizes (r = |Z|/√n_nz, where n_nz is the number of non-tied pairs) were large in Cohen's [23] terms, ranging from 0.68 (BI) to 0.81 (PE).

TABLE II. One-Sample Wilcoxon Signed-Rank Tests vs. Midpoint 3.0 (all p < 0.0001)

| Construct | Mean | Mdn | *n_nz* | *Z* | *r (effect)* |
|---|---|---|---|---|---|
| PE | 4.28 | 4.50 | 98 | 8.05 | 0.81 |
| EE | 4.11 | 4.50 | 91 | 7.22 | 0.76 |
| BI | 4.05 | 4.50 | 95 | 6.63 | 0.68 |
| SE | 4.04 | 4.00 | 92 | 7.32 | 0.76 |
| HM | 4.00 | 4.00 | 92 | 6.69 | 0.70 |
| OQ | 4.16 | 4.50 | 92 | 7.40 | 0.77 |

### C. *Inter-Construct Correlations*

Spearman rank-order correlations among the six construct scores were uniformly positive and statistically significant after Holm correction across all 15 pairwise tests (all adjusted $p < 10^{-11}$). Correlations ranged from $\rho = 0.63$ (PE–SE) to $\rho = 0.77$ (HM–OQ); the median was approximately 0.71, and bootstrap 95% lower-CI bounds ran between 0.48 and 0.66. Such uniformly strong inter-correlations are consistent with the nomological structure predicted by TAM/UTAUT [2]–[5], but warrant a direct empirical check of dimensionality, presented next.

### D. *Dimensionality and Common-Method Bias*

Dimensionality diagnostics on the 12×12 polychoric correlation matrix yielded meritorious sampling adequacy (KMO = 0.847) and rejected the identity hypothesis (Bartlett's $\chi^2(66) = 1385.0$, $p < 0.001$). Horn's parallel analysis [20] retained a single factor: the first polychoric eigenvalue (8.77) exceeded its 95th-percentile random benchmark ($\approx$ 1.76), whereas the second eigenvalue (0.73) did not. Harman's single-factor test [16] yielded a first-factor variance of 73.1%, exceeding the 50% rule-of-thumb threshold used to flag common-method bias; a one-factor extraction produced standardized loadings of 0.79–0.90 on every item. Two interpretations follow. Substantively, students' responses to a short, post-workshop instrument organize around a single general acceptance attitude rather than six empirically separable facets; the six constructs function as interpretable sub-dimensions but are not cleanly discriminable in this sample. Methodologically, the large first factor is consistent with common-method bias from self-report at a single time point; any inter-construct correlation should therefore be read as an upper-bound estimate of the true covariance.

### E. *Between-Session Comparison*

Kruskal–Wallis tests across the three workshops (n = 31 / 45 / 27) were non-significant for every construct (PE $H = 0.26$, $p = 0.88$; EE $H = 1.83$, $p = 0.40$; BI $H = 0.21$, $p = 0.90$; SE $H = 1.60$, $p = 0.45$; HM $H = 0.36$, $p = 0.84$; OQ $H = 0.03$, $p = 0.99$) and for the 12-item grand mean ($H = 0.008$, $p = 0.996$). Construct medians were within 0.50 scale-points across sessions. The three workshops therefore produced statistically indistinguishable acceptance profiles, supporting protocol fidelity and ruling out session-level confounds.

### F. *Respondent Segmentation*

Segmenting respondents by their grand mean across all 12 items using the primary threshold (High ≥ 4.00; Moderate 3.00–3.99; Low < 3.00) yielded 64 High acceptors (62.14%), 31 Moderate (30.10%), and 8 Low (7.77%); the grand mean was M = 4.10 (SD = 0.85, 95% CI [3.94, 4.26]; range 1.42–5.00). Sensitivity analysis over alternate cut-points showed the Low-acceptor proportion to be highly stable (6.80–7.77%) and the High-acceptor proportion to vary with threshold; even under the most conservative cut-point (≥ 4.50), approximately two-fifths remained classified as High acceptors.

### G. *Qualitative Themes from Open-Ended Feedback*

Of 103 responses, 19 rows contained any non-empty free-text entry; five comprised a single dash and one a numeric string (likely a student identifier entered in the wrong field); thirteen substantive comments were retained for thematic analysis. Inductive coding yielded seven themes (counts in parentheses): enthusiasm/enjoyment (6), usefulness for work/future career (2), learning-curve concerns (2), accuracy/reliability concerns (2), integration with conventional workflows (1), need for external assistance (1), and pedagogical suggestions on workshop placement (1); two comments were cross-coded across themes. Representative quotations include "building n8n workflow is quite enjoyable and intuitive" (enthusiasm) and "it's too inconsistent and the prompts yields inaccurate results" [sic] (accuracy/reliability). Because single-coder analysis limits inter-rater credibility, theme counts are reported as illustrative.

## V. Discussion

### A. *Acceptance Profile of AI-Automation Tooling (RQ1)*

RQ1 asked to what extent computer engineering undergraduates accept AI-automation tooling across the six TAM/UTAUT constructs and whether their acceptance is reliably above the neutral midpoint of the scale. The Results section reports that all six construct means and their bootstrap confidence intervals lie above the neutral midpoint with large non-parametric effect sizes. Three substantive readings of this finding are supported by prior literature. First, the result is consistent with the favourable-acceptance baseline that Davis [2] and Venkatesh et al. [3], [4] document for utility-oriented technologies, particularly for users whose daily activities already entail engagement with the technological category in question—in this case, students who routinely use generative AI chat tools and are therefore disposed to view AI-automation tooling as a familiar, lower-risk extension. Second, the absence of a sceptical center of gravity is theoretically meaningful: the UTAUT2 framework [4] predicts that prior exposure and habit raise behavioural intention by reducing perceived novelty cost, a mechanism that fits an undergraduate cohort already embedded in AI-supported coursework. Third, the small but well-defined minority that resists this favourable consensus invites a separate explanation: Hoff and Bashir [9] argue that learned trust attitudes toward automation are shaped by prior exposure to failure and can persist even in environments designed to be trust-supportive—an account consistent with the minority pattern in our data. Taken together, RQ1's answer is that AI-automation tooling enters the educational setting on a favourable footing for the majority of students, but with a non-trivial and theoretically expected minority whose learned-distrust responses warrant pedagogical attention rather than dismissal.

### B. *Construct Inter-relations and Dimensional Collapse (RQ2)*

RQ2 concerned the relative ordering of the six TAM/UTAUT constructs, the strength of their interrelations, and—crucially—whether they are empirically separable. The Results section establishes a tight ordering of construct means with Performance Expectancy at the top and Hedonic Motivation at the bottom, a fully positive and uniformly strong correlation matrix, and a single-factor solution under both Horn's parallel analysis [20] and Harman's diagnostic [16].

The interpretive question is what this empirical collapse signifies.

Two complementary explanations are supported by the literature. Substantively, in a controlled instructional setting where the technology is novel and exposure is short, students do not yet have the differentiated experiential basis required to discriminate among acceptance facets that the model treats as conceptually distinct. Davis's original TAM derivation [2] was based on multi-week organizational use, and the longitudinal extension in TAM2 [5] explicitly notes that construct discriminability strengthens with cumulative experience. A short post-workshop instrument administered to first-time learners should therefore elicit a more globally evaluative—halo-like—response rather than a fully differentiated cognitive structure. That an instrumental construct (PE) anchors the top of the ranking and an affective construct (HM) the bottom is itself meaningful: students frame AI-automation tooling primarily through a utilitarian lens [2], [5], with intrinsic enjoyment subordinate to utility, in line with Davis, Bagozzi and Warshaw's [14] extrinsic–intrinsic distinction.

Methodologically, single-source self-report instruments inflate inter-construct covariance through common-method bias [16], and short two-item scales further amplify this effect by reducing each construct's unique variance, so the observed correlations should be read as upper-bound estimates of the true covariance. The dimensionality diagnostics cannot mechanically separate the substantive halo from the method-induced halo, which is precisely why both interpretations should be carried forward. Neither undermines the descriptive utility of the construct ordering, but both caution against treating the six constructs as distinct intervention targets in a short instrument—a recommendation aligned with Eisinga, te Grotenhuis and Pelzer [17] on parsimonious scales and Podsakoff et al. [16] on remedies for common-method variance. The implication is that short post-workshop instruments are best read as measures of a single attitudinal dimension supplemented by content-specific qualitative probes.

### C. Quantitative-Qualitative Integration of Student Attitudes (RQ3)

RQ3 asked how the qualitative student narratives relate to the quantitative acceptance profile. The Results section reports four converging patterns and one substantive divergence. Here the divergence carries the most theoretical weight. On the convergent side, the qualitative themes track the quantitative ranking in the way mixed-methods integration would predict [4], [5]: instrumental endorsements correspond to the high-ranking Performance Expectancy construct; expressions of enjoyment cluster among engaged students even though Hedonic Motivation is comparatively low overall, supporting the UTAUT2 view of HM as a sub-population effect rather than a uniform attribute [4]; comments about steep learning curves and uneven instruction sequencing align with the middling Effort Expectancy result and reproduce the construct's classical sensitivity to scaffolding [3]; and student remarks about needing external assistance to build unaided are consistent with Bandura's [6] and Compeau and Higgins's [7] characterization of self-efficacy as a scaffold-dependent attitude that the early stages of mastery have not yet fully resolved. These convergences should be read alongside the upper-bound nature of the inter-construct correlations noted earlier.

The divergence on Output Quality is the more interesting finding because it cannot be read directly from either stream alone. The quantitative score for OQ is high and would, taken in isolation, support a confident conclusion that students view automated outputs as accurate. The qualitative stream complicates this picture: a small subset of substantive comments (n = 2, reported as illustrative rather than representative) describes outputs as inconsistent and error-prone. Read through the trust-in-automation literature, this is exactly the pattern that Lee and See [8] describe as the appropriate–reliance problem, that Parasuraman and Riley [10] map onto the use–misuse–disuse–abuse continuum, and that Hoff and Bashir [9] decompose into dispositional, situational, and learned trust layers. The implication is that aggregate OQ endorsement does not equal calibrated trust. Recent human–AI interaction research operationalizes this insight: Bansal et al. [12] demonstrate that human–AI team performance is shaped by whether users verify recommendations rather than defer to them, and Buçinca, Malaya and Gajos [13] show that cognitive forcing functions can reduce overreliance in AI-assisted decision-making. For an AI-automation platform whose outputs are produced by opaque language-model agents, the pedagogical reading of the OQ divergence is that students require structured opportunities to encounter, recognize, and respond to failure, not merely opportunities to observe success—a position that places verification behaviour, rather than acceptance attitude, at the center of pedagogical design [11]–[13].

Read together, the three RQs describe a population that broadly accepts AI-automation tooling on instrumental grounds, that does not yet differentiate among the conceptual sub-facets of acceptance after a short exposure, and whose attitudinal stance toward automation reliability is a more demanding educational target than the headline endorsement suggests.

### D. Theoretical, Methodological, and Pedagogical Contributions

Empirical and methodological. The study provides construct-level acceptance evidence on Thai computer engineering undergraduates—a population under-represented in the TAM/UTAUT literature—across six constructs of the TAM/UTAUT family, with protocol fidelity verified across three identically scripted sessions. We are not aware of prior acceptance studies of an AI-augmented low-code automation platform [15] in undergraduate computing education in a South-East Asian setting. Methodologically, the paper demonstrates a principled matching of analyses to a parsimonious, ordinal, single-source instrument—polychoric inputs for ordinal items, Spearman–Brown and McDonald's omega for two-item scales [17], [19], bootstrap CIs, Holm correction across pairwise tests [22], polychoric KMO and Bartlett's tests [21], Horn's parallel analysis [20], and Harman's single-factor diagnostic [16]—rather than forcing structural equation modelling on an under-identified design, and is offered as a methodological template for educational-technology acceptance studies operating under similar measurement constraints.

Theoretical and pedagogical. The empirical collapse of the six TAM/UTAUT sub-facets into a single general acceptance attitude in short-form post-workshop measurement is itself a theoretical finding: it suggests caution against treating these constructs as empirically discriminable in studies whose measurement design resembles the present one—short-form, single-occasion, two indicators per construct, n ≈ 100—and prompts a refinement of TAM/UTAUT scope—the full nomological network should be expected to manifest after sustained exposure rather than after a single instructional event [2], [5]. Pedagogically, the mixed-methods integration identifies three concrete instructional levers grounded in trust-in-automation theory [8]–[11] and human–AI interaction research [12], [13]: (i) instruction-sequencing scaffolds for effort-expectancy friction; (ii) self-efficacy supports for the gap between recognizing usefulness and feeling capable; and (iii) explicit reliability demonstrations that turn aggregate output-quality endorsement into calibrated trust. Most consequentially, evaluation of student acceptance should not be reduced to a single immediately-post-instruction survey: longer-horizon, behaviourally anchored measurement is required to capture whether favourable initial attitudes mature into appropriately calibrated reliance [8], [12], [13].

### *E. Limitations, Future Work, and Conclusion*

Several limitations apply. The single-institution, cross-sectional, self-report design precludes causal inference and leaves acquiescence and order effects uncontrolled; two items per construct preclude confirmatory or structural modelling, and Social Influence and Facilitating Conditions are omitted. The Harman result signals non-trivial common-method variance, so inter-construct correlations are upper-bound estimates [16]. Open-ended comments were single-coded with a 48-hour dual-pass check, and session assignment was reconstructed from timestamps, though Kruskal–Wallis tests rule out session-level confounds. Future work should pursue multi-institution longitudinal replication linking intention to actual use logs, expand to ≥ 3 items per construct for confirmatory and discriminant-validity testing with Social Influence and Facilitating Conditions reintroduced, apply temporal separation and a marker-variable design [16], run a randomised Output Quality manipulation (failure-case vs. standard) to test trust calibration [8]–[13], and use dual-coded interviews and platform-comparative studies (e.g., Zapier) to triangulate findings.

Computer engineering undergraduates' acceptance of AI-automation tooling was broadly favourable across all six TAM/UTAUT constructs, anchored on instrumental usefulness rather than intrinsic enjoyment. The qualitative stream converged on usefulness and enthusiasm but diverged on output quality, surfacing a reliability-sceptical minority. Dimensionality diagnostics revealed that the canonical sub-facets are not empirically separable in this short-form post-workshop context, behaving as sub-dimensions of a general acceptance attitude — a methodologically and theoretically consequential finding. The results support curricular adoption and identify three theory-grounded instructional levers — instruction-sequencing, graduated-independence, and trust-calibration — for improvement.

## REFERENCES

[1] n8n GmbH, "n8n: Workflow automation tool," n8n.io. [Online]. Available: https://n8n.io. [Accessed: 24-Apr-2026].

[2] F. D. Davis, "Perceived usefulness, perceived ease of use, and user acceptance of information technology," *MIS Quart.*, vol. 13, no. 3, pp. 319–340, Sep. 1989.

[3] V. Venkatesh, M. G. Morris, G. B. Davis, and F. D. Davis, "User acceptance of information technology: Toward a unified view," *MIS Quart.*, vol. 27, no. 3, pp. 425–478, Sep. 2003.

[4] V. Venkatesh, J. Y. L. Thong, and X. Xu, "Consumer acceptance and use of information technology: Extending the unified theory of acceptance and use of technology," *MIS Quart.*, vol. 36, no. 1, pp. 157–178, Mar. 2012.

[5] V. Venkatesh and F. D. Davis, "A theoretical extension of the technology acceptance model: Four longitudinal field studies," *Manage. Sci.*, vol. 46, no. 2, pp. 186–204, Feb. 2000.

[6] A. Bandura, "Self-efficacy: Toward a unifying theory of behavioral change," *Psychol. Rev.*, vol. 84, no. 2, pp. 191–215, Mar. 1977.

[7] D. R. Compeau and C. A. Higgins, "Computer self-efficacy: Development of a measure and initial test," *MIS Quart.*, vol. 19, no. 2, pp. 189–211, Jun. 1995.

[8] J. D. Lee and K. A. See, "Trust in automation: Designing for appropriate reliance," *Human Factors*, vol. 46, no. 1, pp. 50–80, Mar. 2004.

[9] K. A. Hoff and M. Bashir, "Trust in automation: Integrating empirical evidence on factors that influence trust," *Human Factors*, vol. 57, no. 3, pp. 407–434, May 2015.

[10] R. Parasuraman and V. Riley, "Humans and automation: Use, misuse, disuse, abuse," *Human Factors*, vol. 39, no. 2, pp. 230–253, Jun. 1997.

[11] R. Parasuraman, T. B. Sheridan, and C. D. Wickens, "A model for types and levels of human interaction with automation," *IEEE Trans. Syst., Man, Cybern. A, Syst. Humans*, vol. 30, no. 3, pp. 286–297, May 2000.

[12] G. Bansal, T. Wu, J. Zhou, R. Fok, B. Nushi, E. Kamar, M. T. Ribeiro, and D. S. Weld, "Does the whole exceed its parts? The effect of AI explanations on complementary team performance," in *Proc. 2021 CHI Conf. Human Factors Comput. Syst. (CHI '21)*, Yokohama, Japan, 2021, Art. 81, pp. 1–16.

[13] Z. Buçinca, M. B. Malaya, and K. Z. Gajos, "To trust or to think: Cognitive forcing functions can reduce overreliance on AI in AI-assisted decision-making," *Proc. ACM Hum.-Comput. Interact.*, vol. 5, no. CSCW1, Art. 188, pp. 1–21, Apr. 2021.

[14] F. D. Davis, R. P. Bagozzi, and P. R. Warshaw, "Extrinsic and intrinsic motivation to use computers in the workplace," *J. Appl. Soc. Psychol.*, vol. 22, no. 14, pp. 1111–1132, Jul. 1992.

[15] A. Sahay, A. Indamutsa, D. Di Ruscio, and A. Pierantonio, "Supporting the understanding and comparison of low-code development platforms," in *Proc. 46th Euromicro Conf. Softw. Eng. Adv. Appl. (SEAA)*, Portorož, Slovenia, 2020, pp. 171–178.

[16] P. M. Podsakoff, S. B. MacKenzie, J.-Y. Lee, and N. P. Podsakoff, "Common method biases in behavioral research: A critical review of the literature and recommended remedies," *J. Appl. Psychol.*, vol. 88, no. 5, pp. 879–903, Oct. 2003.

[17] R. Eisinga, M. te Grotenhuis, and B. Pelzer, "The reliability of a two-item scale: Pearson, Cronbach, or Spearman–Brown?" *Int. J. Public Health*, vol. 58, no. 4, pp. 637–642, Aug. 2013.

[18] L. J. Cronbach, "Coefficient alpha and the internal structure of tests," *Psychometrika*, vol. 16, no. 3, pp. 297–334, Sep. 1951.

[19] R. P. McDonald, *Test Theory: A Unified Treatment*. Mahwah, NJ, USA: Lawrence Erlbaum, 1999.

[20] J. L. Horn, "A rationale and test for the number of factors in factor analysis," *Psychometrika*, vol. 30, no. 2, pp. 179–185, Jun. 1965.

[21] H. F. Kaiser and J. Rice, "Little Jiffy, Mark IV," *Educ. Psychol. Meas.*, vol. 34, no. 1, pp. 111–117, Apr. 1974.

[22] S. Holm, "A simple sequentially rejective multiple test procedure," *Scand. J. Stat.*, vol. 6, no. 2, pp. 65–70, 1979.

[23] J. Cohen, "A power primer," *Psychol. Bull.*, vol. 112, no. 1, pp. 155–159, Jul. 1992.

[24] H. N. Boone and D. A. Boone, "Analyzing Likert data," *J. Extension*, vol. 50, no. 2, Art. 48, Apr. 2012, doi: 10.34068/joe.50.02.48.

[25] V. Braun and V. Clarke, "Using thematic analysis in psychology," *Qual. Res. Psychol.*, vol. 3, no. 2, pp. 77–101, 2006.